\documentstyle[twoside,fleqn,espcrc2,epsf]{article}

\title{Non-Universality Effects and Dark Matter in Gravity Mediated SUSY Breaking}

\author{R. Arnowitt
\address{Center for Theoretical Physics, Department of Physics,
Texas A\&M University,\\
College Station, TX 77843-4242} 
and Pran Nath
\address{Department of Physics, Northeastern University,
Boston, MA  02115}}
    
\begin{document}

\begin{abstract}
Dark matter detection rates for supergravity models with R parity where supersymmetry is
broken at a scale $\stackrel{>}{\sim} M_G$ are discussed.  Non-universal soft breaking
masses in both the Higgs and squark sectors are considered, and it is seen that these can
effect rates by a factor of 10 - 100 when
$m_{\tilde\chi_{1}^0}\stackrel{<}{\sim}$ 65 GeV ($\chi_1^0$ = lightest
neutralino) but otherwise make relatively small corrections.  The
$b\rightarrow s+\gamma$ branching ratio is seen to correlate with
detector event rates, large (small) branching ratios corresponding
to small (large) event rates.  Effects of precision
determinations of cosmological parameters or event rate
predictions by future satellite experiments are discussed for the
$\Lambda$CDM and the $\nu$CDM models \end{abstract}

\maketitle
\noindent
{\bf 1.~~INTRODUCTION}\\

\indent
The dark matter problem is a particularly interesting one to examine in
supersymmetry since models with R-parity invariance automatically have a dark
matter candidate, the lightest supersymmetric particle (LSP).  Over most of the
parameter space in supergravity models, the LSP is the lightest neutralino
$(\chi_1^0$), and so predictions for event rates for detection can be made.

We consider here models based on supergravity, where SUSY is broken in a hidden
sector at a scale $\stackrel{>}{\sim} M_G$ (where $M_G\cong 2\times 10^{16}$ Gev
is the GUT scale) with this breaking communicated to the physical sector by
gravity [1], and with radiative breaking of SU(2) x U(1) occurring at the
electroweak scale $\sim M_Z$ [2].  The simplest such model is the minimal one
[MSGM] with universal soft breaking at $M_G$ [1,3].  Such models depend on only
four additional parameters and one sign.  These can be taken to be $m_0$ (the
universal scalar soft breaking mass), $m_{1/2}$ (the universal scalar soft
breaking mass), $A_0$ (the universal cubic soft breaking parameter), $B_0$ (the
quadratic soft breaking parameter) and the sign of $\mu_0$ (where $\mu_0$ is the
Higgs mixing parameter in the superpotential, $W_{\mu} = \mu_0~H_1H_2$).  After
SU(2) x U(1) breaking, one may alternately replace $B_0$ by tan$\beta =\langle
H_2\rangle/\langle H_2\rangle$ ($H_{1,2}$ are the two Higgs doublets of SUSY),
$m_{1/2}$ by the gluino mass ($m_{\tilde g}\cong
(\alpha_3(M_Z)/\alpha_G) m_{1/2}$, where $\alpha_G\cong 1/24$ is the GUT coupling
constant) and $A_0$ by $A_t$ (the t-quark A parameter at the electroweak scale).

Until recently, almost all calculations on detection of dark matter (DM) has
been done within this universal framework.  However, one expects the
possibility of non-universalities arising from the following sources:  (1)
Kahler potential interactions can give rise to non-universal soft breaking. 
(2) Even if universality holds at a higher scale e.g. the string scale
$M_{str}$, the running of the renormalization group equations (RGE) will
generate non-universalities at $M_G$.  (3) In breaking higher rank grand
unified groups to the Standard Model (SM) group, the D terms can generate
non-universalities.

The number of parameters needed to describe the non-universality expected at
$M_G$ depends upon the gauge group.  For example, for SU(5), neglecting phases
and splittings in the first two generations (to supress flavor changing neutral
currents) one needs nine additional observable parameters, of which only four
enter significantly for large segments of the parameter space.  (This might be
compared with the $\approx$ 30 parameters of the MSSM).  Over the past two
years effort has been made to explore this larger parameter space [4,5], and we
will describe here some results that have been obtained.  We will see that
non-universal masses can occur both in the Higgs and sfermions sectors, and
these two sectors can effect each other either constructively or destructively.

There are several phenomena which effect predicted dark matter rates, and
we summarize these now:

\begin{enumerate}
\item
t-quark mass ($m_t\cong$ 175 GeV).  The heavy top drives the lightest top
squark, $\tilde t_1$, tachyonic for negative $A_t$, eliminating this part of
the parameter space unless $A_t/m_0\stackrel{>}{\sim}$ -0.5.
\item
$b\rightarrow s+\gamma$ decay.  The current CLEO branching ratio is
$B[B\rightarrow X_s\gamma$] = (2.32$\pm$ 0.67) $\times\\
10^{-4}$ [6] which
can be compared with the SM calculation (with NLO corrections) of
$B[b\rightarrow s\gamma$] = (3.28 $\pm$ 0.33) $\times$ 10$^{-4}$ [7].  It is
clear that already the experimental value constrains any new physics
corrections that raise the theoretical value, and one finds that most of
the $A_t/\mu <0$ region is already eliminated at the 95\% C.L.  Combined with the
t-quark effect above, most of the $\mu < 0$ part of the parameter space has been
eliminated [5]. 
\item
Amount of cold dark matter (CDM).  The astronomical determinations of the
various cosmological parameters at present have\\
large uncertainties.  Thus the
Hubble constant, H = h (100 km/s $Mpc$, has the range of 0.5 $\stackrel
{<}{\sim}$ h $\stackrel {<}{\sim}$ 0.75.  The density of matter $\rho_i$ of
type {\it i} can be measured by $\Omega_i =\rho_i/\rho_c$ where the critial
density $\rho_c$ is given by $\rho_c = 3H^2/8\pi G_N\cong 1.88~\times
10^{-29}h^2$ gm/cm$^3$.  The determinations for cold dark matter
(non-relativistic matter at the time of galaxy formation) are in the
range 0.3 $\stackrel{<}{\sim} \Omega_{CDM}\stackrel{<}{\sim}$ 0.75.  Thus
one has 
\end{enumerate}
\begin{equation}
0.1\leq\Omega_{CDM} h^2\leq 0.4
\end{equation}

\noindent
We will assume here that this CDM are the relic neutralinos $\chi_1^0$.  Eq.
(1) represents the cosmological abundance of $\chi_1^0$.  Terrestial detectors
can observe the CDM in the Milky Way impinging on the solar system.  This local
density of $\chi_1^0$ has uncertainties due to modeling of the halo of the
Galaxy and the amount of machos in the Galaxy.  We will assume here that the
local density is $\rho_{\chi_{1}}$ = 0.3 GeV/cm$^3$, though this number could
be in error by a factor of 2 or more.

The calculation of predicted event rates then proceeds as follows:  one first
calculates the relic density given by [8]

$$
\Omega_{\chi_{1}^0} h^2 \cong 2.45 \times 10^{-11}
\biggl(\frac{T_{\chi_{1}^0}}{T_{\gamma}}\biggr)^3
\biggl(\frac{T_{\gamma}}{2.73^o}\biggr)^3\nonumber\\
$$ 
$$
~~~~~~~~~~~~~~~\times N_f^{1/2}/J(x_f)~~~~~~~~~~~~~~~~~~~~~~~~~~~~~{(2)}
$$

\noindent
where $J(x_f)$=$\int_0^{x_{f}}$ dx $< \sigma v >$ GeV$^{-2}$,
$x_f=kT_f/m_{\chi_{1}^0}$, $T_f$ is the freezeout temperature, $N_f$
the number of degrees of freedom at freezeout, $\sigma$ is the $\chi_1^0$
annihilation cross section, $v$ is the relative velocity, and $< >$ means
thermal average.  We restrict the SUSY parameter space such that
$\Omega_{\chi_{1}^{0}}h^2$ falls within the allowed window of Eq. (1), and also
that the SUSY bounds from LEP, Tevatron and CLEO be obeyed.  Within this
restricted parameter space, one then calculates the detector event rate R
given by [9]

$$
R=(R_{SI} +R_{SD}) (\rho_{\chi_{1}^{0}}/0.3 GeV cm^{-3})\nonumber\\
$$
$$
~~~~~~~~~~~~~~(v_{\chi_{1}^{0}}/320 kms^{-1})
~~\frac{events}{kg~d}~~~~~~~~~~~~~~{(3)}  $$

\medskip
\noindent
where $R_{SI} = 16 m_{\chi_{1}^{0}}M_N^3M_Z^4[M_N+m_{\chi_{1}^{0}}]^{-2}\mid
A_{SI}\mid^2$ and $R_{SD}=16 m_{\chi_{1}^{0}}
M_N[M_N+m_{\chi_{1}^{0}}]^{-2}\lambda^2 J(J+1)\mid A_{SD}\mid^2$.  Where $M_N$
is the target nuclear mass and J is its spin, and $A_{SI}$, $A_{SD}$ are the spin
independent, spin dependent scattering amplitudes.  Note that for large $M_N$ one
has $R_{SI}\sim M_N$, while $R_{SD}\sim 1/M_N$ thus favoring the spin independent
scattering with heavy nuclei.

\bigskip
\noindent
{\bf 2.~~$\mu^2$ Dependence on Soft Breaking Masses}
\medskip

\indent
We will assume here that the first two generations of sfermions are
degenerate (to suppress flavor changing neutral currents) and allow for
non-universal soft breaking masses in the Higgs and third generation
sfermions.  We parameterize these at $M_G$ as follows:
\setcounter{equation}{3}
\begin{equation}
m_{H_{1}^2} = m_0^2(1+\delta_1);~~
m_{H_{2}^2} = m_0^2(1+\delta_2)~~~~~
\end{equation}

\begin{eqnarray}
m_{q_{L}}^2 &=& m_0^2(1+\delta_3);~~
m_{u_{R}}^2 = m_0^2 (1+\delta_4);\nonumber\\
m_{d_{R}}^2 &=& m_0^2 (1+\delta_5)
\end{eqnarray}

\begin{equation}
m_{d_{L}}^2 = m_0^2 (1+\delta_6);~~
m_{\ell_{L}}^2 = m_0^2 (1+\delta_7)
\end{equation}

\noindent
Here $q_L = (u_L, d_L)$ are the doublet of squarks, $\ell_L =
(\nu_L, e_L)$ the doublet of sleptons and the reference mass $m_0$ is
taken to be the common mass of the first two generations.  In addition
there are the t, b and $\tau$ cubic soft breaking parameters $A_{ot}$,
$A_{ob}$, $A_{o\tau}$.  For grand unified models with GUT groups
containing an SU(5) subgroup (e.g. SU(N), $N\geq 5$; SO(N), $N\geq 10$,
$E_6$ etc.) and with matter in the usual 10 + $\overline 5$ of SU(5),
one has
\begin{equation}
\delta_3 = \delta_4 = \delta_5;~~\delta_6 = \delta_7;~~
A_{ob} = A_{o\tau}
\end{equation} 

In the following, we will limit our parameter space so that $m_0$,
$m_{\tilde g}\leq$ 1 TeV, tan$\beta\leq$ 25 and  -1 $\leq \delta_i\leq
1$.  For tan$\beta$ in this domain, results are generally insensitive
to $\delta_5,~\delta_6,~\delta_7$ $A_{ob}$ and $A_{o\tau}$, and we will
set these parameters to zero.  The radiative breaking of SU(2) x U(1)
determines $\mu^2$ to be 

\begin{equation}
\mu^2 = \frac{\mu_1^2 - \mu_2^2 t^2}{t^2-1} - \frac{1}{2} M_Z^2;~~t\equiv
tan\beta
\end{equation}

\noindent
where $\mu_i^2 = m_{H_{i}}^2 +\Sigma_i$ and $m_{H_{i}}^2$ are the
running Higgs masses, and $\Sigma_i$ are loop corrections.  One finds
[5]

$$
\mu^2 =
\frac{t^2}{t^2-1}\biggl[\biggl(\frac{1-3
D_0}{2}+\frac{1}{t^2}\biggr)\nonumber\\
$$
$$
+\biggl([\delta_3+\delta_4]\frac{1-D_0}{2}
-\delta_2
\frac{1+D_0}{2}\nonumber\\
$$
$$
+\frac{\delta_1}{t^2}\biggr)\biggr]~m_0^2
+\frac{t^2}{t^2-1}\nonumber\\
$$
$$
\biggl[\frac{1}{2}(1-D_0)\frac{A_R^2}{D_0}
+ C m_{\tilde g}^2\biggr]\nonumber\\
$$
$$ 
- \frac{1}{2} M_Z^2 +\frac{1}{22}
\frac{t^2+1}{t^2-1}
(1-\frac{\alpha_1}{\alpha_G}) S_0\nonumber\\
$$
$$
~~~~~~~~~~~+~loop~corrections~~~~~~~~~~~~~~~~~~~~~~{(9)}
$$
\setcounter{equation}{9}
\noindent
where $D_0=1-({m_t}/200 sin\beta)^2$, $A_R\cong A_t$ - 0.61 $m_{\tilde
g}$, $S_0=TrYm^2$, Y is the hypercharge and $m^2$ in $S_0$ are the scalar
masses at $M_G$.  $D_0$ vanishes at the t-quark Landau pole, and in general
is small ($D_0\stackrel{<}{\sim}$ 0.25).  $A_R$ is the residue at the
Landau pole.
\smallskip
One sees several features of Eq. (9).  For $t^2 >> 1$ (i.e.
t$\stackrel{>}{\sim}$ 3) $\delta_1$ does not enter sensitively in
$\mu^2$.  Further, since $D_0$ is small, we see that $\delta_3$ and
$\delta_4$ acts oppositely to $\delta_2$, i.e. the squark
non-universalities can contribute constructively or distructively to the
Higgs non-universality.  One thus cannot consider only one type of
non-universality.

\begin{figure}
\begin{center}
\mbox{\epsfxsize=3.0in \epsffile{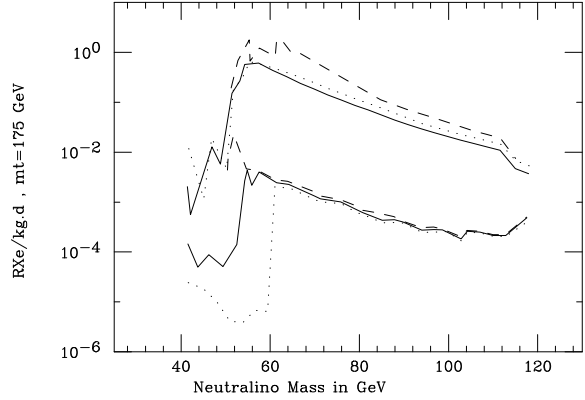}}
\end{center}
\caption{Maximum and minimum event rates for a xenon
detector for universal soft breaking (solid), $\delta_2 =-1=-\delta_1$
(dotted) and $\delta_2=1=-\delta_2$ (dashed) [5].}
\end{figure}

In general, one can expect important modifications due to non-universal
soft breaking to occur when, for some reason, the universal terms are
small.  This can occur if $D_0\cong \frac{1}{3}$, or when the residue at
the Landau pole vanishes $(A_t\cong 0.6 m_{\tilde g})$, or if $m_{\tilde
g}$ is small (i.e. if $m_{\chi_{1}^0}$ is
small since for much of the parameter space $\mu^2/M_Z^2 >> 1$ and one
has the scaling $m_{\chi_{1}}^0\simeq (1/7) m_{\tilde g}$ [10]).  These
effects are enhanced for small $tan~\beta$.  We illustrate some of
these effects for DM detection event rates.  $\mu^2$ gives rise to small
event rates and small $\mu^2$ plays a key role here in that it governs the
interference between the Higgsino and gaugino parts of the $\chi_1^0$ in the SI
part of the $\chi_1^0$ - nuclear scattering cross section.  In general, large
$\mu^2$ gives rise to small event rates and small $\mu^2$ to large event rates
[11].  Fig. 1 shows the maximum and minimum event rates for a xenon detector
for universal and non-universal soft breaking masses.  Here
$\delta_3=0=\delta_4$.  One sees the non-universal effects are small for large
neutralino masses ($m_{\chi_{1}}^0\stackrel{>}{\sim}$ 60 GeV).  For
$\delta_2=-1=-\delta_1$, there can be a reduction of a factor $\sim$ 10 - 100
in the minimum event rates (where tan$\beta$ is small) for small
$m_{\chi_{1}^0}$, since then by Eq. (9) $\mu^2$ is increased by the
non-universalities.  Correspondingly, for $\delta_2=1=-\delta_2$, $\mu^2$ is
decreased and event rates can be increased by a factor $\sim$ 10.  Fig. 2 shows
the corresponding curves for $\delta_1=\delta_2=\delta_3=0$.  One sees that the
$\delta_4=+1$ dotted curve, resembles the $\delta_2=-1$ curve of Fig. 1, and
the $\delta_4=-1$ dashed curve resembles the $\delta_2=+1$ curve of Fig. 1, as
one would expect from Eq. (9).

\begin{figure}
\begin{center}
\mbox{\epsfxsize=3.0in \epsffile{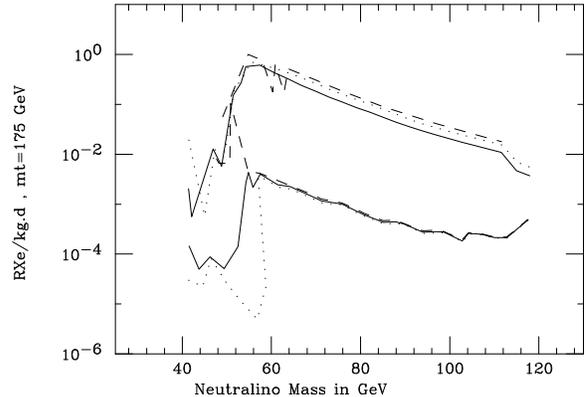} }
\end{center}
\caption{
Maximum and minimum event rates for a xenon detector for
universal softbreaking (solid), $\delta_4=+1$ (dotted), and
$\delta_4=-1$ (dashed) [5].}
\end{figure}

\noindent
Fig. 3 exhibits the fact that when $\delta_3=1=\delta_4$ (as expected in GUT models) and
$\delta_2=1$, the squark and Higgs non-universal effects act coherrently
to significantly increase the maximum event rates up to $\sim$ 10
event/kg d (which is the current level of dark matter detector
sensitivity [4]).  However, for $\delta_2=-1=-\delta_1$, the two effects
mostly cancel, yielding event rates close to predictions of universal
soft breaking.

\begin{figure}
\begin{center}
\mbox{\epsfxsize=3.0in \epsffile{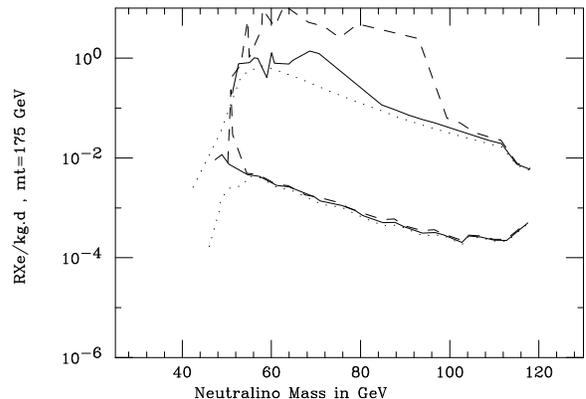} }
\end{center}
\caption{Maximum and minimum event rates for a xenon detector for
universal soft breaking (solid), $\delta_3=1=\delta_4$,
$\delta_2=-1=-\delta_1$ (dotted), and $\delta_3=1=\delta_4$,
$\delta_2=1=-\delta_2$ (dashed) [5].}
\end{figure}

\medskip

\noindent
{\bf 3.~~Cosmological Parameters}

\smallskip

As discussed in Sec. 1, the various cosmological parameters are at
present not well determined.  However, future satellite experiments, MAP
and PLANCK [12] will be able to measure the angular power spectrum quite
accurately, and this will allow the determination of the Hubble constant,
the amount of dark matter, the cosmological constant etc. at the level
of (1-10)\% [13,14].  Such determinations would considerably restrict the
SUSY paramter space, and hence sharpen significantly the predictions of
DM detection rates.  To illustrate what might be expected from these
determinations, we consider two cosmological models.

\medskip
\noindent
(i)~~$\Lambda$CDM Model\\

\indent
One assumes there that CDM and baryonic dark matter (B) exists with a
cosmological constant $\Lambda$ such that the universe is flat: 
$\Omega_{CDM}+\Omega_B+\Omega_{\Lambda}$ = 1.  As an example, we assume that
the measured central values of the parameters are $\Omega_{CDM}$ = 0.40,
$\Omega_B$ = 0.05, $\Omega_{\Lambda}$ = 0.55 and h = 0.62, which are consistent
with current astronomical measurements.  Then using the estimated accuracy that
could be achieved by the PLANCK satellite [13], one finds
\setcounter{equation}{9}
\begin{equation}
\Omega_{CDM} h^2 = 0.154\pm 0.017
\end{equation}

\noindent
This window is much narrower then what is currently assumed, i.e. Eq. (1). 
Eq. (10) produces two interesting results for the event rates:  the
minimum event rates are significantly raised for
$m_{\chi_{1}}^0\stackrel{>}{\sim}$ 60 GeV, and the upper bound on
$\Omega_{\chi_{1}}^0 h^2$ produces an upper bound on allowed values of 
$m_{\chi_{1}}^0$.  One finds at the 1$\sigma$ (2$\sigma$) ranges that
$m_{\chi_{1}}^0\leq$ 70 (77) GeV, and by scaling this produces a corresponding
bound $m_{\tilde g}\leq$ 520 (560) GeV.  In addition
$m_{\chi_1^{\pm}}\stackrel{<}{\sim}$ 150 GeV and for the light Higgs one has
$m_h\stackrel{<}{\sim}$ 120 GeV.  It is interesting to compare these results with
the reach of the upgraded Tevatron with 25 $fb^{-1}$ of data where the gluino
would be observable if $m_{\tilde g}\stackrel{<}{\sim}$ 450 GeV [15], the chargino if
$m_{\tilde\chi_{1}^{\pm}}\stackrel{<}{\sim}$ 235 GeV for about 2/3 of the
parameter space and the Higgs if $m_h\stackrel{<}{\sim}$ 120 GeV [16].

As discussed in Sec. 1, the $b\rightarrow s+\gamma$ decay branching ratio plays
an important role in limiting the SUSY parameter space, as there is already
some strain between the experimental branching ratio and the SM prediction. 
In addition, there exists an interesting correlation between the DM event rate R
and B($b\rightarrow s+\gamma$).  Large R occurs mainly for large tan$\beta$
where destructive interference between the SM and SUSY contributions to the
$b\rightarrow s+\gamma$ occurs.  Small R occurs mainly at small tan$\beta$ where
there is constructive interference in the $b\rightarrow s+\gamma$ decay and
hence correlates with large B ($b\rightarrow s+\gamma$).  This can be seen in
Fig. 4 where a scatter plot for R vs. B$(b\rightarrow s+\gamma$) over the SUSY
parameter space is shown for the $\Lambda$CDM model (assuming universal soft
breaking masses).  One sees for almost all parameter points that, if
B$(b\rightarrow s+\gamma) > 3\times 10^{-4}$ (the 1 std. lower bound of the
SM), then $R <$ 0.1 events/kg d, while if B$(b\rightarrow s+\gamma) < 3 \times
10^{-4}$, then $R > $ 0.05 events /kg d.  Thus new CLEO data on the
$b\rightarrow s+\gamma$ decay will have significant effects on predictions of
dark matter event rates.

\begin{figure}
\begin{center}
\mbox{\epsfxsize=3.0in \epsffile{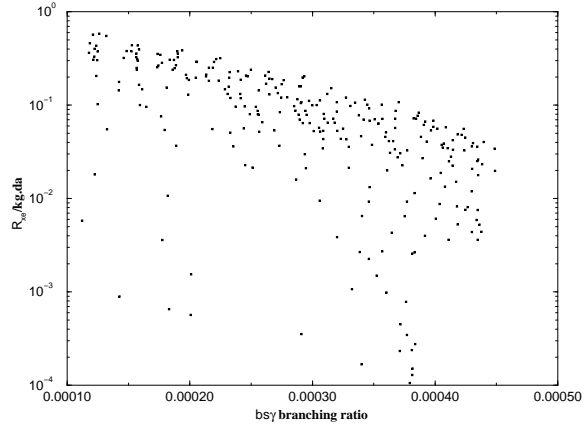} }
\end{center}
\caption{Scatter plot of R vs. B($b\rightarrow s+\gamma$) for a Xe detector for
the $\Lambda$CDM model with universal soft breaking masses and $\mu > 0$.}
\end{figure}
\medskip

\noindent
(ii) $\nu$CDM Model
\smallskip

If neutralinos have mass of order of a few eV, they could represent a hot dark matter
component to the dark matter.  As an example of such a model, we assume that the measured
central values of the cosmological parameters are
$$
\Omega_{\nu}= 0.20;~~\Omega_{CDM} = 0.75,\nonumber\\
$$
$$
~~~~~~~~~~~~~~~~~\Omega_B= 0.05, ~~h = 0.62~~~~~~~~~~~~~~~~~~{(11)}
$$
\medskip
\noindent
The PLANCK satellite would then determine\hfill\\
$\Omega_{CDM}h^2$ with the following
accuracy [13,14]:
\setcounter{equation}{11}
\begin{equation}
\Omega_{CDM}h^2 = 0.288\pm 0.013
\end{equation}
\noindent 
For this case, the narrowing of the $\Omega h^2$ window,
narrows the separation between maximum and minimum event rates in the region
$m_{\chi_{1}}\stackrel{<}{\sim}$ 65 GeV, and for the 1std range (0.158 $\leq
\Omega_{\chi_{1}}h^2\leq$ 0.301) produces forbidden gaps when $m_{\chi_{1}} > $
65 GeV (which, however, get filled in for the 2 std range).  Since
$\Omega_{\chi_{1}} h^2$ is larger for this case then in the $\Lambda$CDM
model, the upper bounds on the gaugino masses are larger.  One finds for the 1
std (2 std) bounds that $m_{\chi_{1}^0}$ $\leq$ 95 (100) GeV, $m_{\tilde g}
\leq$ 700 (720) GeV and for the chargino $m_{\chi_{1}^{\pm}}\stackrel{<}{\sim}$
200 GeV.  Thus for this model, the LHC would most likely be needed to discover
SUSY.

The above two examples are meant to be illustrative of what future astronomical measurements
of the basic cosmological parameters will be able to achieve.  In particular one sees how
astronomical measurements can impact on accelerator searches for SUSY particles.

\medskip
\noindent
{\bf 4.~~Conclusions}
\smallskip

We have considered here dark matter detection rates within the framework of supergravity
grand unification with R-parity where SUSY breaks at a scale $\stackrel{>}{\sim} M_G$, the
breaking being communicated to the physical sector by gravity.  Such models automatically
imply the existance of cold dark matter, and over a large amount of the
parameter space in amounts consistent with astronomical measurements.

The detection rates expected for cases of non-universal soft breaking masses have been
compared with the minimal universal SUGRA models.  Non-universal effects can increase or
decrease event rates (depending on the signs of the non-universalities) by factors of
$\sim$ 10 - 100 in the domain $m_{\chi_{1}}^0\stackrel{<}{\sim}$ 65 GeV (equivalently
when $m_{\tilde g}\stackrel{<}{\sim}$ 400 GeV) but generally have small effects at higher
masses.  One must also consider the possibilities of both Higgs and squark mass
non-universalities, as both these enter in the analysis with comparable size.  Thus the
Higgs and squark non-universalities can either cancel or enhance each other in the
detection rates.

Future $b\rightarrow s +\gamma$ decay data may play an  important role in uncovering new
physics, and indeed may be the first place that new physics is seen.  Already the current
data strongly restricts the SUSY parameter space.  Thus combined the fact that
the top quark is heavy ($m_t\cong$ 175 GeV) the current branching ratio for
$b\rightarrow s+\gamma$ forbids most of the $\mu < 0$ part of the parameter
space, eliminating most of the high event rate region for $\mu < 0$.  The event
rates for $\mu < 0$ are then $\sim$ 100 times smaller than for $\mu > 0$, and
this phenomena holds with or without non-universal soft breaking masses.  As
error flags go down, the $b\rightarrow s+\gamma$ decay may more strongly
restrict the allowed regions of SUSY parameter space, making supergravity
predictions more precise.

One of the striking features of recent years is the ``astro-particle connection'' where
develop-\hfill\\
ments in particle theory effect astronomical and cosmological
theory, and astronomical measurements can influence what is expected at high
energy accelerators.  One may expect this interaction to streng\-then in the
future.  Thus new satellites (MAP and PLANCK), balloon and ground based
experiments should be able to measure the basic cosmological parameters very
accurately (at the few percent level), and hence determine the amounts of
different types of dark matter.  This will put further restrictions on the
allowed SUSY parameter space.  In this connection we have considered two
examples of cosmological models, the $\Lambda$CDM model (which limit the
gaugino masses to be $m_{\chi_{1}^0}\stackrel{<}{\sim}$ 75 GeV, $m_{\tilde
g}\stackrel{<}{\sim}$ 550 GeV and $m_{\chi_{1}^{\pm}}\stackrel{<}{\sim}$ 150 GeV,
and the $\nu$CDM model (where $m_{\chi_{1}^0}\stackrel{<}{\sim}$ 100 GeV, 
$m_{\tilde g}\stackrel{<}{\sim}$ 700 GeV and
$m_{\chi_{1}^{\pm}}\stackrel{<}{\sim}$ 200 GeV.  In the first model the upgraded
Tevatron might be able to see SUSY, while in the second model one likely needs
the LHC (or NLC).  In either case the astronomical measurements would correlate
with accelerator searches for SUSY.  There is also a correlation betweeen the
B($b\rightarrow s+\gamma$) branching ratio and expected dark matter detection
rates.  Thus for $B > 3 \times 10^{-4}$ (i.e. B in range expected from the SM)
the dark matter detector event rates are generally small i.e. $R\simeq
(10^{-1}-10^{-4})$  events/kg d, while if $B < 3 \times 10^{-4}$ (smaller than
the SM prediction) event rates in general will be large i.e. $R\simeq (0.05 -
1)$ events/kg d.  Thus accelerator measurements would correlate with
astronomical searches for dark matter.

\medskip
\noindent
{\bf 5.~~Acknowledgements}\\

This work was supported in part by NSF grants PHY-9411543 and PHY-9602074.\\

\medskip
\noindent
{\bf 6.~~References}
\begin{enumerate}
\item
A.H. Chamseddine, R. Arnowitt and P. Nath, Phys. Rev. Lett. {\bf 49}, 970
(1982).  For reviews see P. Nath, R. Arnowitt and A.H. Chamseddine, ``Applied N
= 1 Supergravity'' (World Scientific, Singapore, 1984); H.P. Nille, Phys. Rep.
{\bf 100}, 1 (1984); R. Arnowitt and P. Nath, Proc. of VII J.A. Swieca Summer
School ed. E. Eboli (World Scientific, Singapore, 1994).
\item
K. Inoue et al. Prog. Theor. Phys. {\bf 68}, 927 (1982); L. Iba$\tilde n$ez and
G.G. Ross, Phys. Lett. {\bf B110}, 227 (1982); L. Avarez-Gaum\'e, J. Polchinski
and M.B. Wise, Nucl. Phys. {\bf B221}, 495 (1983), J. Ellis, J. Hagelin, D.V.
Nanopoulos and K. Tamvakis, Phys. Lett. {\bf B125}. 2275 (1983); L.E.
Iba${\tilde n}$ez and C. Lopez, Nucl. Phys. {\bf B233}, 545 (1984).
\item
L. Hall, J. Lykken and S. Weinberg, Phys. Rev. {\bf D27}, 2359 (1983); P. Nath, R. Arnowitt
and A.H. Chamseddine, Nucl. Phys. {bf B227}, 121 (1983). 
\item
V. Berezinsky, A. Bottino, J. Ellis, N. Forrengo, G. Mignola, and S. Scopel,
Astropart. Phys. 5:1 (1996); ibid, 5:  333 (1996).
\item
P. Nath and R. Arnowitt, hep-ph/9701301.
\item
 M.S. Alam et al. (CLEO Collaboration), Phys. Rev. Lett. {\bf 74}, 2885 (1995).
\item
K. Chetyrkin, M. Misiak and M. Munz, hep-ph/96 12313.
\item
B.W. Lee and S. Weinberg, Phys. Rev. Lett. {\bf 39}, 165 (1977); D.A. Dicus, E. Kolb and V.
Teplitz, Phys. Rev. Lett. {\bf 39}, 168 (1977); H. Goldberg, Phys. Rev. Lett.
{\bf 50}, 1419 (1983); J. Ellis, J.S. Hagelin, D.V. Nanopoulos and N.
Srednicki, Nucl. Phys. {\bf B238}, 453 (1984); J. Lopez, D.V. Nanopoulos and K.
Yuan, Nucl. Phys. {\bf B370} 445 (1992); M. Drees and M.M. Nojiri, Phys. Rev.
{\bf D47},. 376 (1993). \item
M.W. Goodman and E. Witten, Phys. Rev. {\bf D31}, 3059 (1983); K. Greist, Phys. Rev. {\bf
D38}, 2357 (1988); {\bf D39}, 3802 (1989)(E); J. Ellis and R. Flores, Phys. Lett. {\bf
B300}, 175 (1993); R. Barbieri, M. Frigeni and G.F. Giudice, Nucl. Phys. {\bf
B313}, 725 (1989); M. Srednicki and R. Watkins, Phys. Lett. {\bf B225}, 140
(1989); R. Flores, K. Olive and M. Srednicki, Phys. Lett. {\bf B237}, 72
(1990).   \item R. Arnowitt and P. Nath,
Phys. Rev. Lett. {\bf 69}, 725 (1992); P. Nath and R. Arnowitt, Phys. Lett. {\bf B289}, 368
(1992). 
\item R. Arnowitt and P. Nath, Phys. Rev. {\bf D54}, 237t (1996).
\item
http://map.gsfc.nasa.gov/\\
http://astro.estec.esa.n$\ell$/SA-general/\\ 
Projects/Cobras/cobras.htm.
\item A. Kosowsky, M. Kamionkowski, G. Jungman and D. Spergel,
Nucl. Phys. Proc. Suppl. {\bf 51B}, 49 (1996).
\item
S. Dodelson, E. Gates and A. Stebbins, Astroph. J. {\bf 467}, 10 (1996).
\item
T. Kamon, J.L. Lopez, P. McIntyre and J.T. White, Phys. Rev. {\bf D50}, 5676 (1994). 
\item
Report of the tev-2000 Study Group, eds. D. Amidei and R. Brock,
FERMILAB-Pub-961082. 

\end{enumerate}
\end{document}